\documentstyle[12pt,aps]{revtex} 
\input{psfig.sty}

\begin{document}
\title{Universal correlations in vibrational spectra of complex crystals}
\author{G. Fagas, Vladimir I. Fal'ko, C.J. Lambert}
\address{School of Physics and Chemistry, Lancaster University, LA1 4YB, UK}
\maketitle

\begin{abstract}
We study spectral statistics of lattice modes in a disordered crystal and in
a crystal with a complex unit cell. The correlations of the
eigenmode frequencies of a block of a disordered solid is found to obey the GOE
Wigner-Dyson statistics. In contrast, the set of eigenfrequencies of a
crystal with a complex unit cell taken at the same point ${\bf Q}%
\not=0$ in the Brillouin zone exhibit correlations specific to the
GUE universality class.
\end{abstract}

\newpage

The Wigner and Dyson statistical approach is extensively used to
analyse the spectra of complex dynamical systems \cite{Wigner}. It has
been proven theoretically and confirmed experimentally that the quantum
energy levels in objects with chaotic dynamics universally obey the same
mutual correlations as the eigenvalues of random symmetric, Hermitian or
symplectic matrices \cite{Mehta}. Although an original statistical analysis
of level correlation was focused at the spectra of microscopic
quantum objects, such as nuclei and molecules \cite{Wigner}, a similar idea
may be applied to any spectral problem related to random matrices, for
example, to the spectra of electromagnetic or electron waves in mesoscopic
cavities \cite{Simons}. Below, we report the results of a statistical
analysis of vibrational spectra of a solid where complexity and chaos in the
phonon spectrum are introduced via a random distribution of masses of the nuclei.

{\it A priori}, two spectral problems can be formulated for a complex solid.
One is about statistical properties of resonances in the absorption spectra of
a finite-size crystalline specimen, which can be regarded as
a very big molecular cluster. A similar problem has been investigated
experimentally \cite{Ellegaard} in chaotic solid resonators where the spectra reveal
correlations specific to Gaussian symmetric real random matrices (the
Gaussian orthogonal ensemble - GOE \cite{Mehta}), which is
reproduced by numerical simulations reported below. The other problem is
related to the spectrum of vibrations in a crystal with a unit cell composed of
many different atoms - that is, of the optical phonons -
taken at a point in the Brillouin zone of the complex solid. One can measure
such spectra in the inelastic scattering experiments. We show that the latter
spectra obey the statistics and correlation laws of the eigenvalues of Hermitian
Gaussian random matrices (unitary ensemble, GUE \cite{Mehta}), which are
usually attributed to the energy spectra of electrons in chaotic cavities in
a magnetic field \cite{Simons}. 

As a model of a chaotic acoustic resonator, we simulate numerically an
fcc-crystal consisted of $12\times 10\times 8$ atoms along the $[010],[101]$,%
$[\bar{1}01]$ cubic crystallographic directions, with random masses of sites
and identical pair interactions between atoms. The latter choice allows us
to avoid complications of a search of an appropriate ground state of the
system. Each configuration of site masses, $\{m_{\bf j}\}$ is produced by a
random number generator. The latter is characterized by an amplitude of a
fluctuation $\left| \delta m\right| /\langle m\rangle =0.3$ around the mean value
$\langle m\rangle $. For $\left| \delta m\right| /\langle m\rangle \ll 1$,
this is the model of a solid mixture of isotopes. The Debye
frequency, $\omega _{D}\approx 2.9\sqrt{k/\left\langle m\right\rangle }$ is
determined by $\langle m\rangle $ and the interatomic force constant $k$,
and it sets us a unit to express the values of eigenfrequencies. Below, we
treat such a model in the harmonic approximation. The dynamics of a lattice
is described by the linearized equations of motion for atomic displacements $%
{\bf u_{j}}$ with respect to the equilibrium positions ${\bf j}$ in the
lattice. In the nearest-neighbor approximation, $m_{{\bf j}}\ddot{u}_{%
{\bf j}}^{\alpha }=-\sum_{{\bf i}}K_{{\bf ij}}^{\alpha \beta }u_{{\bf i}%
}^{\beta }$, where $K_{{\bf ij}}^{\alpha \beta }=k(l_{{\bf ij}}^{\alpha }l_{%
{\bf ij}}^{\beta }-4\delta _{{\bf ij}}\delta ^{\alpha \beta })$, $l_{{\bf ij}%
}^{\alpha }={\bf (j-i)}^{\alpha }/{\bf \left| j-i\right| }$, and ${\bf (j-i)}$
is taken from the first coordination sphere. We find the set of
eigenfrequencies, $\{\omega _{n}\}$, by solving numerically the eigenvalue
problem

\begin{equation}
det(D-\omega^{2}\;I)=0\;,\;D=M^{-1/2}KM^{-1/2},  \label{eq2}
\end{equation}
where the randomness is brought in by a random diagonal matrix $M_{{\bf ij}%
}^{\alpha \beta }=m_{{\bf i}}\delta ^{\alpha \beta }\delta _{{\bf ij}}$. The
dynamical matrix $D$ is diagonalized by a standard NAG library black box
routine for each given distribution of masses over the lattice.

Repeated periodically, the same block of atoms represents the model of a
solid with a complex unit cell. The spectrum of optical lattice vibrations
of such an object may be derived independently for each value of wavevector $%
{\bf Q}$ within the reduced Brillouin zone, whereas the box size plays now the
role of a complex lattice period. The spectrum of optical modes
can be found by solving Eq. (\ref{eq2}) with a modified dynamical matrix
which incorporates the boundary condition on the edges of a unit cell
imposed by the finite wave-number induced phase shift in the lattice
displacements:

\[
K_{{\bf ij}}^{\alpha \beta }({\bf Q})=\sum_{{\bf h}}k(l_{{\bf h+i,j}%
}^{\alpha }l_{{\bf h+i,j}}^{\beta }\;\ e^{\imath {\bf Qh}}-4\delta _{{\bf %
h+i,j}}\delta ^{\alpha \beta }),
\]
where ${\bf h}=L{\bf h}_{1}+M{\bf h}_{2}+N{\bf h}_{3}$ belongs to the
Bravais lattice of a complex crystal and ${\bf i,j}$'s are the atomic
positions within the unit cell. Note that for ${\bf Q\neq 0}$, the matrix
elements of $D({\bf Q})$ related to the sites on the edges of the unit cell
acquire complex phase factors, which make the whole dynamical matrix
complex,

\[
D({\bf Q})=D_{S}({\bf Q})+\imath D_{A}({\bf Q}),
\]
where $D_{S}({\bf Q})$, $D_{A}({\bf Q})$ are real symmetric and
antisymmetric matrices, respectively. In the particular cases of ${\bf Q}=0$%
, or ${\bf Q}$ taken in the corners of the Brillouin zone of a complex
crystal, such as ${\bf Q}=(0,\pi /|{\bf h}_{2}|,\pi /|{\bf h}_{3}|)$, $D_{A}(%
{\bf Q})=0$ and $D({\bf Q})=D^{S}({\bf Q})$ is real symmetric, as in the
case of acoustic resonances in an isolated block. \ 

The numerically calculated spectrum of modes $\{\omega _{n}\}$ enables us to
derive the distribution function $P(s)$ of the nearest-level-spacing, $%
s=(\omega _{n+1}-\omega _{n})/\Delta $. It is natural to measure spacings in units
of the mean level spacing $\Delta =1/\nu (\omega )$, where $\nu (\omega )$
is the calculated disorder-average density of states. The function $P(s)$ is
built upon a two-step averaging procedure. The first one is to use an
ensemble of 50 random realizations of the distribution of masses in the
crystal. A further averaging can be applied after considering ergodicity in
the chaotic scattering regime: we average $P(s)$ over a broad frequency
range, namely, $\omega \in \lbrack 0.35, 1]$, for each of the calculated
spectra by using an observation that they become self-similar after having
been rescaled by $\Delta $.

The nearest-level-spacing distribution function for a disordered solid
resonator is indistinguishable from that for the optical phonon spectrum at $%
{\bf Q}=0$, and it is shown in Fig. 1(a). It coincides with the random
matrix theory prediction for real symmetric matrices given by the
Wigner-Dyson distribution function for the GOE(plotted with dashed line).

\begin{figure}
\centerline{\psfig{figure=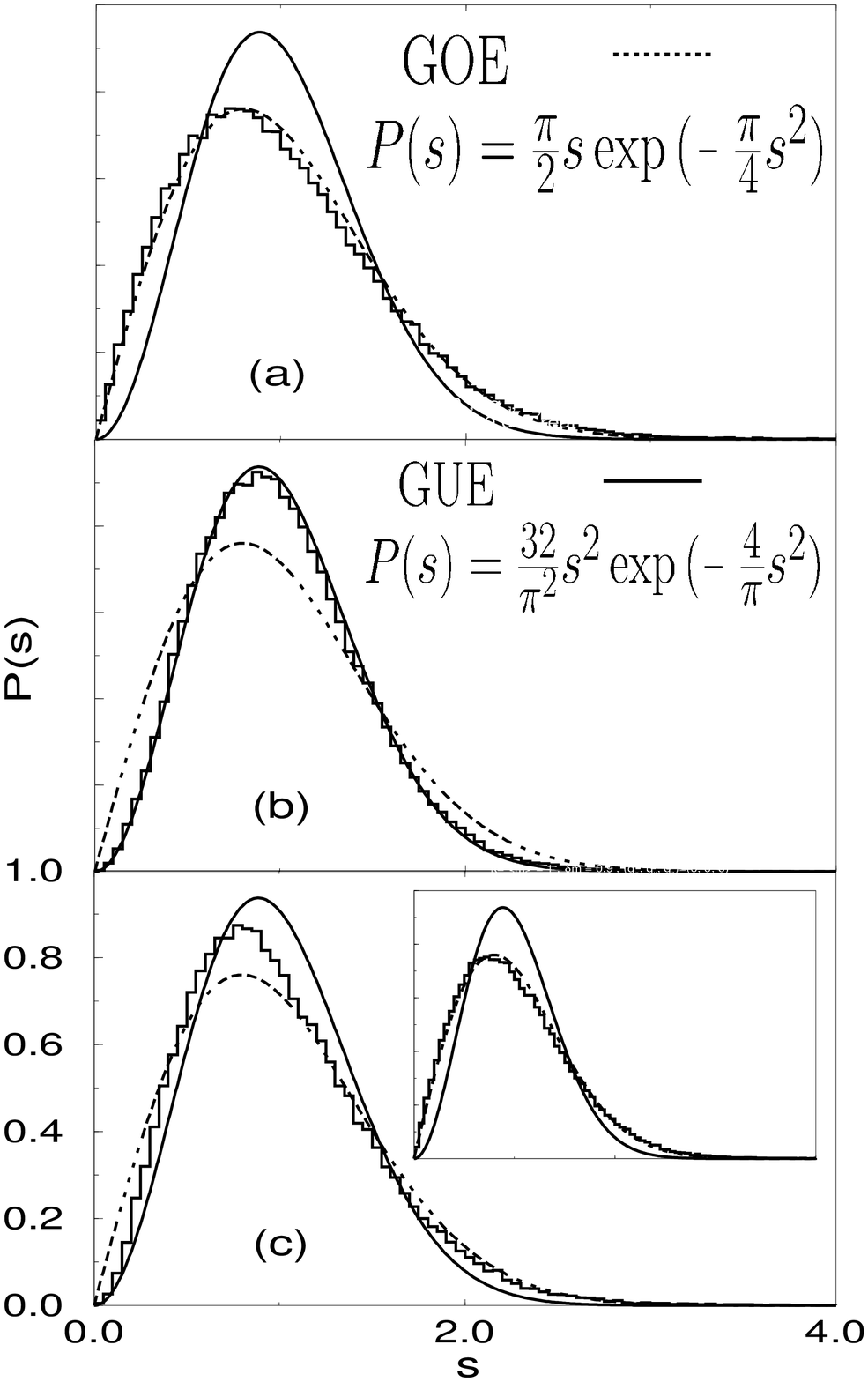,height=12cm,width=7.5cm}}
\caption{$P(s)$ histograms for {\bf Q} equal to (a)$0$,
(b)$[0, \pi/4|{\bf h}_{2}|, 3\pi/4|{\bf h}_{3}|]$,
(c)$[0, 0.1/|{\bf h}_{2}|, 0]$
(inset : ${\bf Q}=[0, \pi/|{\bf h}_{2}|, \pi/|{\bf h}_{3}|]$).}
\end{figure}

In contrast, optical lattice vibrations with wavevector ${\bf Q}\not=0$ show
the nearest-level-spacing statistics which is best fitted by the Wigner-Dyson
distribution function for random Hermitian matrices (GUE). A typical
numerically found $P(s)$-histogram is plotted in Fig. 1(b), in comparison
with the GUE analytical result \cite{Mehta} quoted in the inset. A similar
observation has been recently made about chaotic electronic band structures
for a finite Bloch momentum \cite{Mucciolo}. As a test, the inset in Fig.
1(c) shows the $P(s)$-histogram for the corner of the Brillouin zone, which
should be of a GOE-type, since $D(\left[ 0,\pi /|{\bf h}_{2}|,\pi /|{\bf h}%
_{3}|\right] )$ is real. 

The evolution of the correlation function $P(s)$ as a function of the wave
number is illustrated by one example in Fig. 1(c). It is apparent from this
plot that the crossover between two distinct limits - from GOE to GUE -
takes place at relatively small values of $|{\bf Q}|$ (which is the
parameter responsible for the imaginary part of a dynamical matrix). A high
sensitivity of correlations in the spectrum of optical phonons in a complex
crystal to the rise of an imaginary part of $D$ is in agreement with what is
known about the crossover between GOE and GUE symmetry classes in chaotic
electronic billiards \cite{Efetov}, and it seems that\ it cannot be reduced
to a trivial mixing of two typical (GOE and GUE) distribution functions \cite
{Altland,VF1,VF2}.

\end{document}